\documentclass[twocolumn,noshowpacs,preprintnumbers,amsmath,amssymb,aps]{revtex4-2}
\usepackage [pdftex]{graphicx}  
\usepackage [utf8x] {inputenc}
\usepackage{textcomp}
\usepackage[usenames]{color}
\usepackage{colortbl}
\graphicspath{{pictures/}}
\DeclareGraphicsExtensions{.pdf,.png,.jpg}
\usepackage{indentfirst}
\usepackage[version=3]{mhchem} 
\usepackage{upgreek}

\usepackage{hyperref}

\usepackage{xcolor}
\usepackage{bm}
\usepackage{amsmath}
\usepackage{amsfonts}
\usepackage{amssymb}%
\usepackage{epstopdf}

\usepackage{xr}
\makeatletter
\newcommand*{\addFileDependency}[1]{
  \typeout{(#1)}
  \@addtofilelist{#1}
  \IfFileExists{#1}{}{\typeout{No file #1.}}
}
\makeatother

\newcommand*{\myexternaldocument}[1]{%
    \externaldocument{#1}%
    \addFileDependency{#1.tex}%
    \addFileDependency{#1.aux}%
}
\myexternaldocument{SI}

\begin{document}

\title{Spatial dynamics and nanomechanical signatures of a van der Waals spin glass}

\author{S.Y. Grebenchuk$^{1,*}$}
\author{T.M. Savilov$^{1,2}$}
\author{Y.W. Ho$^{1,3}$}
\author{A.K. Grebenko$^{3}$}
\author{M. Kravtsov$^{1,2}$}
\author{J.M.~Wolf$^{3}$}
\author{D.A. Litvinov$^{1,2}$}
\author{Z. Yang$^{4}$}
\author{P. Cai$^{4}$}
\author{S. Ramachandran$^{4}$}
\author{R. Duan$^{5}$}
\author{L. Zheng$^{5}$}
\author{B.~\"Ozyilmaz$^{1,2,3}$}
\author{G. Eda$^{3,6}$}
\author{E.J.G. Santos$^{4,7}$}
\author{M. Koperski$^{1,2}$}
\author{K.S. Novoselov$^{1,2}$}
\author{M. \v{S}i\v{s}kins$^{1,8,}$}
\email{Corresponding authors: S.Y. Grebenchuk, sergey.g@u.nus.edu; M. \v{S}i\v{s}kins, m.siskins@soton.ac.uk}

\affiliation{$^1$ Institute for Functional Intelligent Materials, National University of Singapore, 117544 Singapore;}
\affiliation{$^2$ Department of Materials Science and Engineering, National University of Singapore, 117575 Singapore;}
\affiliation{$^3$ Department of Physics, National University of Singapore, Singapore, 117551 Singapore;}
\affiliation{$^4$ Institute for Condensed Matter and Complex Systems, School of Physics and Astronomy, University of Edinburgh, Edinburgh EH9 3FD, United Kingdom;}
\affiliation{$^5$ School of Materials Science and Engineering, Nanyang Technological University, Singapore, 639798 Singapore;}
\affiliation{$^6$ Department of Chemistry, National University of Singapore, Singapore, 117583 Singapore;}
\affiliation{$^{7}$ Donostia International Physics Center 20018, Donostia-San Sebasti\'an, Spain;}
\affiliation{$^8$ School of Physics and Astronomy, University of Southampton, Highfield, Southampton SO17 1BJ, United Kingdom}

\begin{abstract}
Spin glasses exhibit complex non-equilibrium magnetic dynamics, yet the mechanical response associated with their slow, history-dependent evolution at the nanoscale remains elusive. Two-dimensional van der Waals spin glass magnets offer a promising platform for exploring the coupling between this glassy magnetism and mechanical degrees of freedom. Here, we investigate the spatial dynamics of magnetic clusters and their magnetoelastic response in the van der Waals spin glass NbFeTe$_2$, which retains glassy magnetic behaviour in exfoliated flakes as thin as $12$~nm. Cryogenic magnetic force microscopy directly visualises the slow spatial reorganisation of magnetic clusters over several hours, while atomistic spin-dynamics simulations reveal competing short-range magnetic interactions as a microscopic origin of the glassy dynamic, demonstrating characteristic ageing. We further incorporate thin NbFeTe$_2$ flakes into suspended nanomechanical resonators, where the slow relaxation of these magnetic clusters modifies membrane tension through exchange-striction coupling. The resulting resonance frequency shifts exhibit stretched exponential relaxation and pronounced magnetic- and thermal-history dependence, providing nanomechanical signatures of spin glass dynamics. Our findings establish nanomechanical resonators as a platform for probing the non-equilibrium magnetism through their mechanical response, opening opportunities to investigate complex magnetic states and develop hybrid magneto-mechanical technologies.
\end{abstract}

\maketitle

\section*{Introduction}

Spin glasses are a unique class of magnetic systems characterised by randomness, competing interactions, and frustration in their spin configurations. Unlike ferromagnets, where spins align, or paramagnets, where disordered spins remain dynamic, spin glasses undergo a transition into a frozen disordered state at low temperatures, accompanied by slow relaxation, ageing, and history-dependent dynamics~\cite{edwards_theory_1975, edwards_theory_1976, binder1986spin, altieri_introduction_2024}. While theoretical models of spin glasses have become paradigms for complex systems, with broad impact ranging from optimisation problems to neural network architectures~\cite{hertz2018introduction}, their experimental study and functional exploitation in devices remain challenging due to slow dynamics, metastability of the spin glass state, and weak coupling to conventional external probes.

Two-dimensional (2D) van der Waals (vdW) magnetic materials provide a promising route to overcome these limitations. The discovery of long-range magnetic order in atomically thin vdW crystals has expanded the landscape of magnetism, enabling unprecedented control over dimensionality, interfaces, and heterostructure design~\cite{burch_magnetism_2018,Gibertini2019,Gong2019,Mak2019,huang_emergent_2020,kurebayashi_magnetism_2022,Genome22,TANG20231}. Incorporating spin glass behaviour into 2D materials platform is particularly appealing: ultrathin frustrated magnets can be mechanically manipulated, integrated into vdW heterostructures, suspended as membranes, and directly coupled to mechanical degrees of freedom~\cite{112_2020,Jiang2020,CGT_Makars}, opening opportunities to access and exploit non-equilibrium magnetic dynamics in functional devices, such as nanoelectromechanical systems (NEMS).


Signatures of spin glass states have recently been reported in a growing number of vdW materials, including \ce{Fe3GeTe2} with intercalated Fe~\cite{pal2024realization}, alkali-ion intercalated \ce{Cr2Ge2Te6}~\cite{khan2024spin}, mixed-cation \ce{XPS3}~\cite{takano2003magnetic, masubuchi2008phase, goossens2013magnetic, graham2020local}, geometrically lattice-frustrated $\beta$--\ce{MoCl4}~\cite{mcguire2018spin}, amorphous CrGeTe~\cite{wang2023spin}, and heavy-fermion antiferromagnetic CeSiI~\cite{torres2024glassy}. Among these materials, \ce{NbFeTe2} (NFT) is particularly promising due to its strong sensitivity to crystal lattice and stoichiometry. Early studies reported spin glass-like behaviour with coexisting ferromagnetic and antiferromagnetic signatures~\cite{zhang_magnetic_1997}, while other samples exhibited bulk ferromagnetism below $T_c \approx 87$ K~\cite{stepanova_bulk_2024}. Subsequent work linked these differences to the stabilisation of either a monoclinic metallic ferromagnet or an orthorhombic semiconducting spin-glass phase~\cite{wu_tailoring_2024}. 

Here, we characterise the cluster spin glass state in orthorhombic NFT, revealing the spatial dynamics of its magnetic clusters, and utilising their magnetoelastic coupling for nanomechanical readout in membrane resonators. Supported by atomistic spin-dynamics simulations that elucidate the role of competing short-range interactions, we experimentally identify history-dependent magnetisation dynamics in NFT with a distinct slow relaxation behaviour in its exfoliated flakes, which persists even in thin layers. Cryogenic magnetic force microscopy directly images the slow reorganisation of magnetic cluster domains in NFT, whose chaotic rearrangements govern this slow, hour-scale macroscopic relaxation. In contrast to scanning probe and magneto-optical techniques, whose sensitivity decreases with flake thickness, nanomechanical probes exhibit enhanced sensitivity due to larger strain responses in thinner membranes~\cite{112_2020,Jiang2020, Xiao25}. Thus, we incorporate thin NFT into suspended nanoelectromechanical membrane resonators capable of detecting the change in strain~\cite{112_2020,CGT_Makars,208_2022,213_2024} associated with this glassy magnetic domain relaxation dynamics with high sensitivity~\cite{Xiao25}. This strain relaxation renders characteristic time-, temperature- and magnetic field-dependent shifts of the resonance frequency, providing a nanomechanical transduction mechanism for spin glass behaviour. This approach establishes coupling of cluster domain dynamics to nanomechanical membranes as a powerful platform for integrating frustrated magnetism into functional NEMS, bridging fundamental spin glass physics with future reconfigurable and low-power device technologies.

\section*{Results}
\begin{figure*}[ht!]
\begin{center}
\includegraphics[width=1.94\columnwidth]{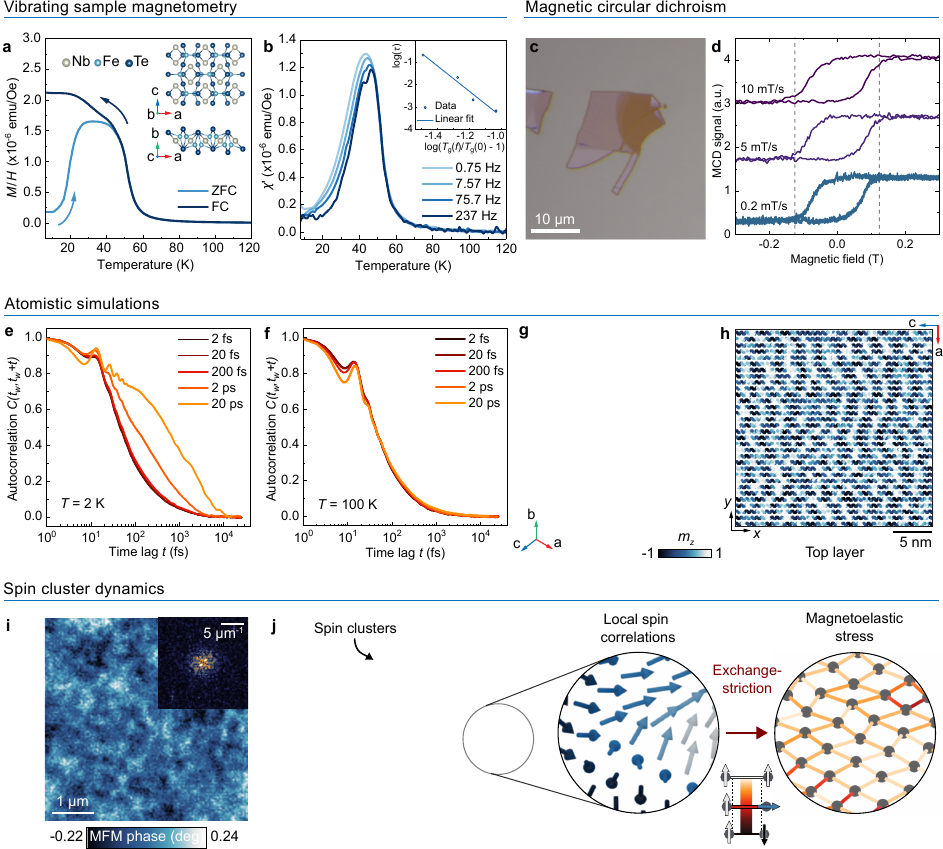}
\caption{\textbf{Spin glass state in NFT.}
\textbf{a} - Vibrating sample magnetometry (VSM) measurements of the temperature dependence of magnetization $M(T)$ normalized by the applied magnetic field $H$ for a bulk crystal under zero-field-cooled (ZFC) and field-cooled (FC) conditions at $H=100$ Oe. Inset: crystal structure of the orthorhombic phase of NFT crystal. 
\textbf{b} - Temperature dependence of real part $\chi'$ of AC magnetic susceptibility $\chi = \chi' + i \chi''$, measured at different frequencies of the AC component of out-of-plane magnetic field. The inset shows fitting of the scaling power law extracted from the glass transition temperature dependence on the frequency of the AC field.
\textbf{c}~-~An optical image of mechanically exfoliated 11~nm thick flake (the darkest color).
\textbf{d} - Dependence of hysteresis curves of 40~nm thick NFT flake on the sweep rate of out-of-plane magnetic field at T = 4~K. 
\textbf{e} - Theoretically calculated spin autocorrelation time lag dependence at different waiting times at T = 2~K.
\textbf{f}~-~Spin autocorrelation time lag dependence at different waiting times above the glass transition temperature, at T = 100~K. Results at other temperatures can be found in the SI in Fig.~S8.
\textbf{g}~-~Atomistic simulation image of a spin configuration after 50 ps relaxation time, featuring formation of spin cluster domains throughout the entire system. The cube size is 25$\times$25$\times$25~nm$^3$. 
\textbf{h} - Top projection of the simulated spin configuration from \textbf{g}.
\textbf{i} - MFM image of 110 nm thick NFT flake at 1.7 K and 0 mT. The inset shows the FFT image, from which the characteristic size of spin clusters is 200-250~nm.
\textbf{j} - Schematic illustration of exchange-striction mediated magnetoelastic coupling in a cluster spin glass. Local spin correlations within and between magnetic clusters can induce magnetoelastic stress through exchange striction. The inset schematically illustrates how the relative orientation of neighbouring spins influences their preferred interatomic separation.
}
\label{fig:1}
\end{center}
\end{figure*}

We first ensure the emergence of spin glass behaviour in NFT by establishing reliable growth of high-quality single crystals and systematically characterising their structural and magnetic properties. Bulk NFT crystals were produced using the chemical vapour transport (CVT) method (see Methods). The resulting crystals have a layered, orthorhombic structure with distinct vdW planes and can be thinned down to several tens of nanometers using conventional mechanical exfoliation methods (Fig.~\ref{fig:1}). Post-growth characterisation of the crystals by Raman spectroscopy, X-ray diffraction and energy dispersive X-ray demonstrates similar results to the LT-phase of NFT observed in prior reports~\cite{wu_tailoring_2024} (see Supplementary Note 1), which is expected to host a spin glass state. 

We further characterise and confirm the magnetic properties of bulk NFT crystals with vibrating sample magnetometry (VSM). Fig.~\ref{fig:1}~\textbf{a} depicts zero field cooling (ZFC) and field cooling (FC) temperature dependencies of the field-normalised magnetisation $M(T)$ of a bulk crystal. $M(T)$ shows distinctive splitting of the ZFC and FC curves as temperature decreases below the glass transition temperature $T_g$~\cite{binder1986spin}, which is estimated to be $T_g\sim 50$ K. This glass transition temperature lies within the reasonable range with previously reported values~\cite{zhang_magnetic_1997, wu_tailoring_2024}, with the slight difference attributed to the expected sensitivity to stoichiometric ratio. We note that $M(H)$ curves for bulk NFT also demonstrate a hysteresis loop with a coercive field around 0.3~T at 2~K, which resembles that of a ferromagnetic material (see Supplementary Note~2), suggesting that the spin glass state in thin layers likely arises from a heavily frustrated ferromagnetism. This frustration of the spins can originate in deviations from the perfect 1-1-2 stoichiometric ratio caused by vacancies or the presence of additional Fe atoms~\cite{wu_tailoring_2024}.

To deduce more information about the nature of spin frustration in NFT, we perform AC magnetic susceptibility measurements at different field modulation frequencies. Temperature dependence of the real part $\chi'$ of the susceptibility (Fig.~\ref{fig:1}~\textbf{b}) demonstrates a shift of the glass transition temperature at small frequencies of the order of 1-100~Hz. This is a direct indication of slow spin dynamics, which is a key feature of a spin glass system. Additional quantitative analysis of the $\chi'(T)$ and $T_g$ dependencies with scaling and power-law~\cite{gunnarsson1988dynamics, mydosh1993spin, benka2022interplay, khan2024spin} provides values of the characteristic relaxation time $\tau_0$ around $7.4\times10^{-10}$~s (inset of Fig.~\ref{fig:1}~\textbf{b}), which corresponds to microscopic flipping for dynamically fluctuating spins or group of spins (see Supplementary Note~3). Our values lie in the range from 10$^{-7}$ to $10^{-11}$~s which is characteristic for cluster spin glass systems~\cite{malinowski2011spin, bag2020study}. 

\subsection*{Slow magnetisation dynamics in thin NFT flakes}

We investigate the persistence of spin glass state with reduced thickness, focusing on whether its characteristic magnetic behaviour is preserved in thin exfoliated NFT flakes. The magnetisation of a single flake is far below the detection limit of conventional VSM, requiring the use of alternative optical probes sensitive to magnetisation. We mechanically exfoliate bulk NFT crystals into thin flakes ranging from $11$ to $110$~nm in thickness (Fig.~\ref{fig:1}~\textbf{c}) and characterise their magnetic properties by magnetic circular dichroism (MCD) measurements (Fig.~\ref{fig:1}~\textbf{d}). The temperature dependence of the hysteresis curves of 40~nm-thin flakes demonstrated similar behaviour to the bulk crystals measured with VSM, with a clear transition temperature around 55~K (see Supplementary Note~5). To confirm the presence of a spin glass state in thin flakes, we measure the hysteresis curves as a function of a magnetic field sweep rate and observe the corresponding change of the coercive field value (Fig.~\ref{fig:1}~\textbf{d}). Conventional ferromagnetic systems react fast towards change of the magnetic field, from pico- to microseconds, making their hysteresis insensitive to a sweep rate within realistic experimental conditions (see Supplementary Note~6), while frustrated systems tend to react significantly slower, up to a few seconds and even hours~\cite{henkel2007ageing, mydosh2015spin}. At faster magnetic-field sweep rates, the spin clusters cannot fully respond to the changing field, requiring higher fields for magnetic reversal and resulting in an increased coercive field. In contrast, slower sweeps allow more time for relaxation, reducing the coercive field. Additionally, a series of time-dependent experiments on responding to changes in the external conditions, such as magnetic field and temperature, revealed slow relaxation dynamics on timescales of several hours (see Supplementary Note 7).

To complement these experimental observations and gain additional microscopic insight, we turn to atomistic spin dynamics methods \cite{Kartsev20,Wahab2021,Zhou25,Xiao25,Haldar25}, which further demonstrate that orthorhombic NFT crystals support the presence of spin glass behaviour with dominant short-range interactions (see Supplementary Note~4). For characterization of the spin glass state, we calculate the spin autocorrelation function $C(t_w,t_w+t) = \langle\mathbf{S_i}(t_w)\cdot\mathbf{S_i}(t_w+t)\rangle$ of the atomic spin $\mathbf{S_i}$ at different temperatures and waiting times $t_\mathbf{w}$ (Fig.~\ref{fig:1}~\textbf{e,f}). At a low temperature of 2~K, $C(t_w,t_w+t)$ autocorrelation functions demonstrate different decay behaviours with several relaxation times, which increase significantly with longer $t_\mathbf{w}$ (Fig.~\ref{fig:1}~\textbf{e}). This behaviour demonstrates the characteristic ageing effect of a glassy state~\cite{kamber_self-induced_2020, verlhac_thermally_2022}. The difference between the relaxation times at different $t_\mathbf{w}$ disappears at higher temperatures with $T > T_g$ (Fig.~\ref{fig:1}~\textbf{f}), with the collapse of all curves at various waiting times to a corresponding paramagnetic state in the system. Further analysis of the spatial distribution maps of the spin configurations shows that spins tend to form clusterous magnetic domains as time evolves since the relaxation starts (Fig.~\ref{fig:1}~\textbf{g,h}). This indicates the presence of strong short-range ferromagnetic correlations, confirming that the cluster-type spin glass state in NFT arises from a heavily frustrated ferromagnetic system. 

\begin{figure*}[ht!]
\begin{center}
\includegraphics[width=1.94\columnwidth]{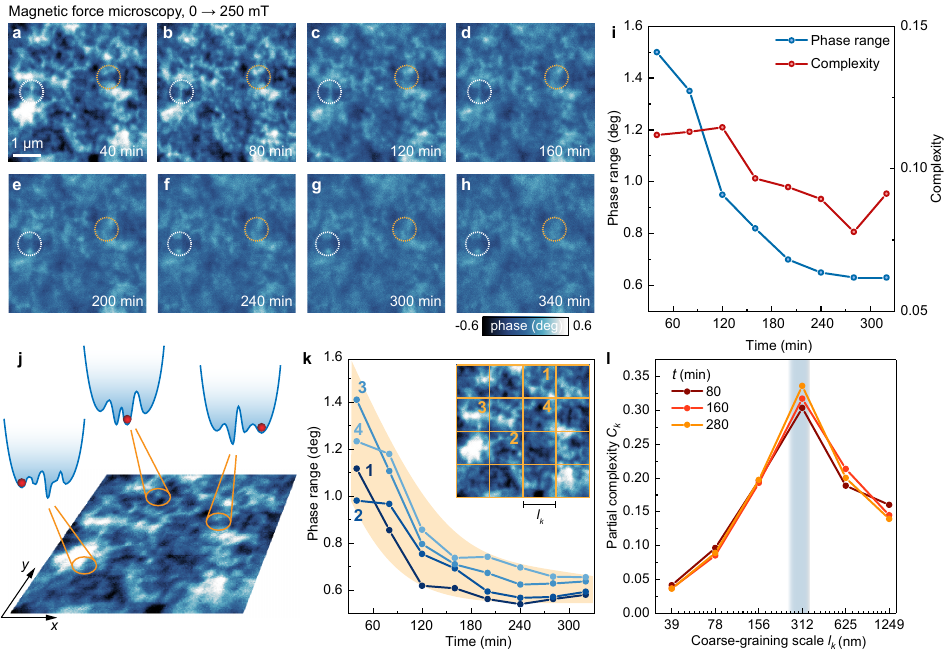}
\caption{\textbf{Magnetic force microscopy visualisation of slow magnetic cluster dynamics.} \textbf{a-h} - Time dependence of the magnetic signal of 110~nm thick NFT flake captured with MFM at 1.7~K and lift height of 30~nm after changing external out-of-plane magnetic field from 0 to 250~mT. Prior to changing the field, the sample stayed in zero field for several hours. \textbf{i} - Phase range (blue curve) and complexity (red curve) time dependencies of MFM images \textbf{a-h}. \textbf{j} - Schematic figure of the spatial distribution of the states at different local minima of the free energy landscape across the sample. \textbf{k} - Examples of phase range time dependence of different regions of the sample shown in the inset figure. 
\textbf{l} - Partial complexity of the normalised phase contrast $C_k$ against the coarse-graining scale $l_k = 2^k \times 39$~nm ($\Lambda = 2$). $C_k = \tfrac{1}{2}\langle(\phi_k - \phi_{k+1})^2\rangle$, where $\phi_k$ is the MFM phase image after $k$ steps of $2\times2$ block averaging, after white noise was removed. The blue vertical line denotes the range of characteristic cluster sizes, estimated from autocorrelation length and from the FFT of MFM images \textbf{(a-h)}.
}
\label{fig:3_MFMvst}
\end{center}
\end{figure*}

The formation of locally correlated spin clusters revealed by atomistic simulations suggests that their slow, collective reorganisation may underlie the long-timescale magnetic relaxation observed experimentally. The spin glass state is usually challenging to directly visualise because of its chaotic nature and need for atomic resolution~\cite{kamber_self-induced_2020, verlhac_thermally_2022,kroeze_directly_2024}. However, we demonstrate that cluster-type spin glass behaviour in NFT allows us to resolve and visualise its magnetisation dynamics on a microscopic scale using cryogenic magnetic force microscopy (MFM). As shown in Fig.~\ref{fig:1}~\textbf{i}, we directly observe that magnetic moments in NFT align to form larger magnetic clusters, which, in contrast to organised domains in ferromagnets, form chaotically disordered patterns.
The local spin correlations within and between these clusters may also influence the lattice through exchange striction, giving rise to magnetoelastic stress and changes in the mechanical properties of NFT (Fig.~\ref{fig:1}~\textbf{j}).

\subsection*{Micro-scale dynamics of magnetic clusters in the spin glass state}

\begin{figure*}[ht!]
\begin{center}
\includegraphics[width=2\columnwidth]{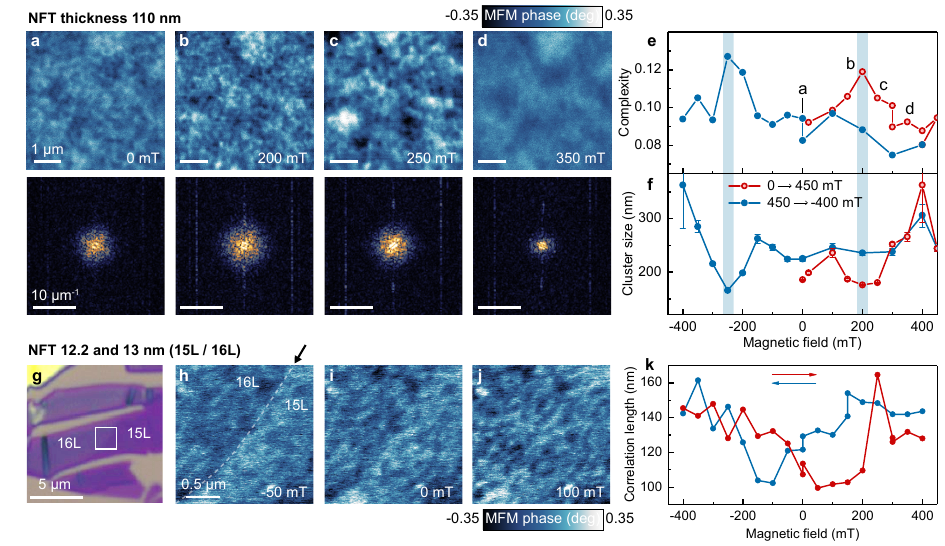}
\caption{\textbf{Magnetic field dependence of magnetic cluster size and structure.} \textbf{a-d} - Top panel: magnetic field dependence of MFM images for a 110~nm thick NFT sample measured at 1.7~K at a lift height of 30~nm. Bottom panel: corresponding FFT images of MFM scans. 
\textbf{e} - Complexity dependence on the external magnetic field extracted from MFM images. 
\textbf{f} - Characteristic cluster size dependence on the magnetic field extracted from FFT spectra of MFM images.
\textbf{g} - Optical image of 12.2 and 13~nm thick NFT sample. 
\textbf{h-j} - Magnetic field dependence of MFM images for the same flake measured at the spot denoted with a white rectangle on panel \textbf{g}. The arrow points out the border between 15L and 16L. MFM was done in an out-of-plane magnetic field at T = 1.7~K and a lift height of 30~nm.
\textbf{k} - Corresponding correlation length dependence on the magnetic field for 15L-16L thin NFT flake.
}
\label{fig:4_MFMvsH}
\end{center}
\end{figure*}

To investigate the slow dynamics of the magnetic clusters identified by MFM, we track their spatial evolution following changes in the applied magnetic field. Fig.~\ref{fig:3_MFMvst}~\textbf{a-h} show a sequence of MFM images acquired after increasing the field from 0 to 250 mT, revealing a gradual decrease in magnetic contrast with time. This relaxation is quantified by the MFM phase range in Fig.~\ref{fig:3_MFMvst}~\textbf{i}, which evolves over several hours until reaching the relaxed state. Alongside the overall reduction in contrast, we observe changes in the spatial arrangement of magnetic clusters. For example, the region highlighted by the white circle shows the gradual disappearance of a narrow magnetic bridge connecting two domains, while at the same time a similar connection emerges in the region marked by the orange circle.

To analyse the dynamics of magnetic microstructure and interactions between groups of spins in detail, we resorted to two main metrics: i) MFM phase range, which is a measure of magnetisation homogeneity across the sample, and ii) multiscale structural complexity~\cite{bagrov2020multiscale}, which characterises the spatial changes in structural composition of magnetic textures (see Methods and Supplementary Note~8). The MFM phase range and complexity are calculated for the set of time-dependent MFM images and plotted in Fig.~\ref{fig:3_MFMvst}~\textbf{i} (see Supplementary Note~8~and~9). Similarly to the phase range, complexity tends to decrease insignificantly over time, albeit non-gradually (the red curve in Fig.~\ref{fig:3_MFMvst}~\textbf{i}). The non-gradual change in complexity can be associated with the glassy state with spatial variations in the magnetic free energy across the sample (Fig.~\ref{fig:3_MFMvst}~\textbf{j}). This behaviour can be understood intuitively by considering the complex magnetic free-energy landscape of a spin glass, characterised by multiple local minima~\cite{mydosh2015spin}, into which different regions of the sample may settle upon cooling. Along with ageing processes, the further relaxation of these states can take different paths, leading to variations in characteristic relaxation times. 

To further investigate this effect, we performed a local analysis of the same images (Fig.~\ref{fig:3_MFMvst}~\textbf{k, l}). Each MFM image was divided into sixteen equal regions, and the evolution of the magnetic contrast in each region was analysed using the same method as in Fig.~\ref{fig:3_MFMvst}~\textbf{i}. Local MFM phase range analysis (Fig.~\ref{fig:3_MFMvst}~\textbf{k}) reveals that the intrinsic state of the system lies in a relatively wide range of values. However, together with the monotonous decrease of its absolute values with time, its MFM phase distribution narrows down as the system goes into a more homogeneous state. 

To determine whether the relaxation preferentially affects particular length scales, we calculated the partial complexity $C_k$ at each coarse-graining step $l_k$~\cite{bagrov2020multiscale} (Fig.~\ref{fig:3_MFMvst}~\textbf{l}). At all waiting times, the partial complexity exhibits a single peak at $l_k=312$~nm, consistent with the correlation length $\xi$ of 280–340 nm. The partial complexity changes little over time, suggesting that while the individual local texture of magnetic clusters relaxes by either merging or separating, the overall structure of the magnetic texture remains largely preserved.

\begin{figure*}[ht!]
\begin{center}
\includegraphics[width=1.9\columnwidth]{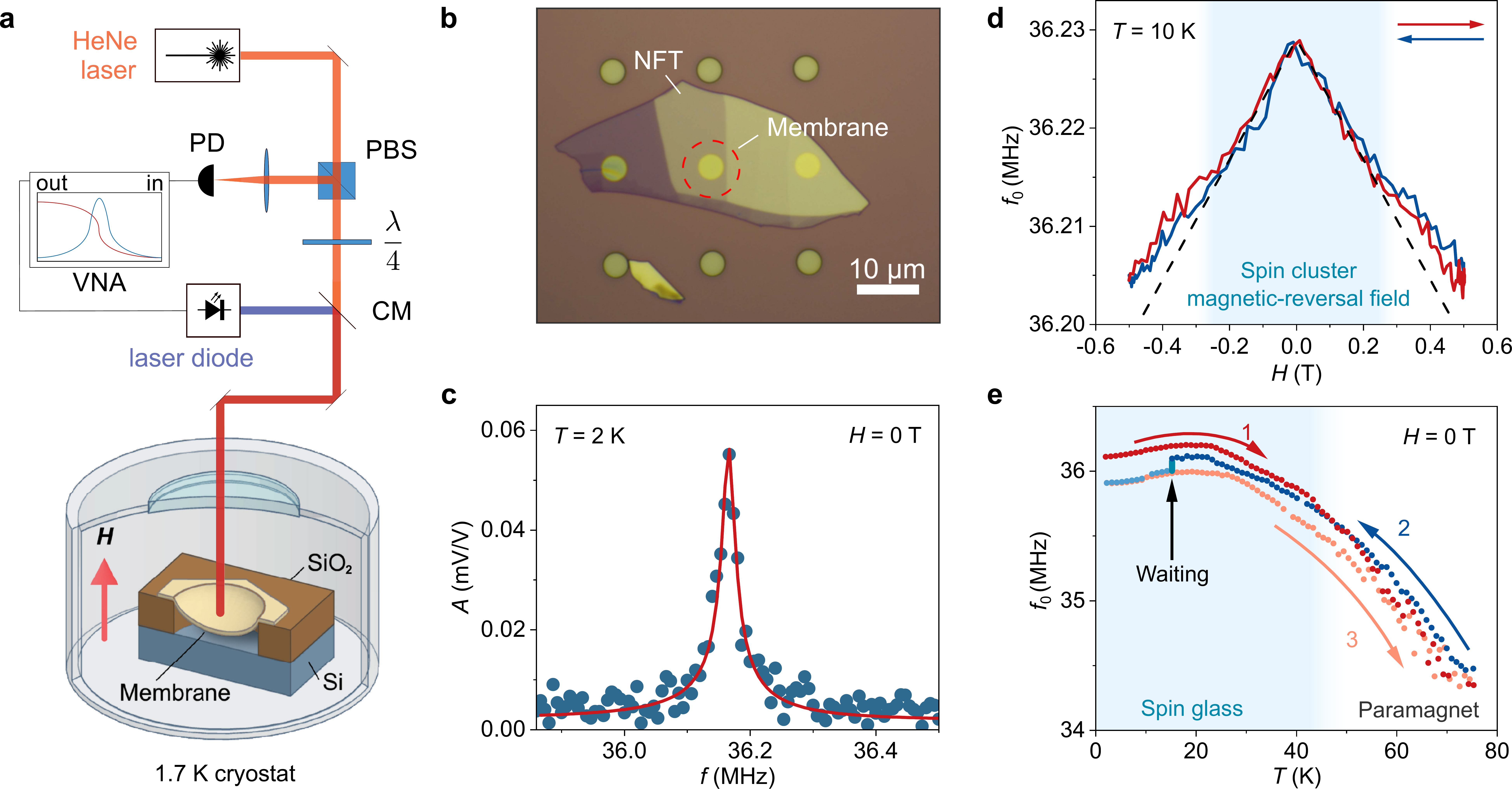}
\caption{\textbf{Coupling glassy magnetism to nanomechanical resonance}. \textbf{a}  – Schematic of the laser interferometry setup, using a $633$ nm laser for detection and a $450$ nm laser for actuation. VNA – vector network analyser; CM – cold mirror; PBS – polarising beam splitter; PD – photodiode. \textbf{b} – Optical image of an NFT membrane suspended over $285$ nm deep circular cavities in SiO$_2$. The measured membrane, outlined by a red dashed circle, has a diameter of $4$ µm and a thickness of approximately $20$ nm. \textbf{c} – Fundamental resonance measured at $2$ K and zero magnetic field (blue dots), fitted with a damped harmonic oscillator model (red line), yielding $f_0=36.165$ MHz and $Q=1980\pm190$. \textbf{d} – Resonance frequency during magnetic field sweeps at $10$ K following $40$ hours of waiting at zero field. The shaded region indicates the spin cluster magnetic-reversal field. \textbf{e} – Temperature dependence of the resonance frequency during a heating–cooling cycle with a stop at $15$ K and subsequent re-heating. Arrows indicate the measurement sequence.}
\label{fig:5}
\end{center}
\end{figure*}

\subsection*{Magnetic field dependence of cluster domains in the
spin glass state}

We further investigate the magnetic field dependence of spin glass clusters observed in the MFM images~(Fig.~\ref{fig:4_MFMvsH}). In contrast to ordinary ferromagnets, where spontaneous magnetisation forms well-defined domains that increase or decrease in size based on the orientation of magnetic moments and the external magnetic field, a frustrated system in NFT exhibits significant spatial reorganisation of its domain clusters, continuously altering its total magnetisation~(top panel in Fig.~\ref{fig:4_MFMvsH}~\textbf{a-d}). Here, the local texture of clusters evolves more chaotically with the magnetic field change, while the overall magnetisation still resembles the $M(H)$ hysteresis curve behaviour (see Supplementary Note~10 for additional MFM data). To determine the characteristic magnetic cluster size distribution, we perform a Fast Fourier Transform (FFT) analysis (bottom panel in Fig.~\ref{fig:4_MFMvsH}~\textbf{a-d}). At 0~mT, disordered clusters are present, giving a circular distribution of sizes in the FFT image~(Fig.\ref{fig:4_MFMvsH}~\textbf{a}). As the magnetic field increases, the number of smaller clusters increases, which is reflected in the broadening of the FFT spot. This broadening corresponds to the emergence of clusters with characteristic sizes ranging from 100 to 200 nm at 200~mT (Fig.~\ref{fig:4_MFMvsH}~\textbf{b}). Upon approaching the magnetic-reversal fields, magnetic moments tend to align in the same direction, leading small clusters to combine more into larger ones, generally increasing the characteristic size to 250$-$300~nm (Fig.~\ref{fig:4_MFMvsH}~\textbf{c}). Eventually, at higher magnetic fields, spins align in one direction, and no fine magnetic microstructure is present (Fig.~\ref{fig:4_MFMvsH}~\textbf{d}). We highlight that the field dependence of the magnetic clusters is also well captured by the complexity metric (Fig.~\ref{fig:4_MFMvsH}~\textbf{e}), which rises at magnetic fields close to coercive force values, when the largest changes in the magnetic textures occur (see Supplementary Note~10). The rise in complexity at coercive field values is also well aligned with spin cluster size reduction (Fig.~\ref{fig:4_MFMvsH}~\textbf{f}) and appearance of finer magnetic structures in the MFM images.

A similar clusterous network is also observed for thinner flakes. Fig.~\ref{fig:4_MFMvsH}~\textbf{g} shows a local spot on the sample with approximately 15 (15L) and 16 layers (16L). The MFM images in Fig.~\ref{fig:4_MFMvsH}~\textbf{h-j} show the magnetic response of the 15L and 16L regions, starting from a negatively magnetised state at $-50$ mT. As the field is increased through zero towards $50-100$ mT, the contrast between the two regions disappears near the magnetic reversal field. At higher fields, the contrast re-emerges, with a stronger signal from the thicker 16L region. Interestingly, the magnetic cluster structure also changes considerably with field, with more pronounced features in thicker flakes, suggesting a thickness dependence of the observed behaviour. The spatial correlation length $\xi$ extracted from the MFM images (see Methods) exhibits a corresponding hysteretic dependence on magnetic field (Fig.~\ref{fig:4_MFMvsH}~\textbf{k}). The minima near $100-150$ mT occur close to the magnetic reversal fields, consistent with the MCD measurements on thin flakes (Fig.~\ref{fig:1}~\textbf{d}), and at lower fields than the characteristic changes observed in the thicker sample (Fig.~\ref{fig:4_MFMvsH}~\textbf{f}).

\subsection*{Nanomechanical coupling to spin-glass cluster domains}
The slow, history-dependent reorganisation of magnetic clusters revealed by MFM provides a pathway for coupling spin glass dynamics to mechanical degrees of freedom. As illustrated schematically in Fig.~\ref{fig:1}~\textbf{j}, changes in the relative orientation and correlations of magnetic moments during cluster reorganisation can modify the equilibrium lattice strain through exchange magnetostriction. When NFT is suspended as a free-standing membrane, these changes can alter its internal tension and thus its mechanical resonance frequency~\cite{112_2020,Jiang2020,CGT_Makars,Xiao25,208_2022,213_2024}. This mechanism offers a route to transduce the collective dynamics of frustrated spins into a measurable nanomechanical response.

To investigate this coupling, we incorporate thin NFT flakes into suspended nanomechanical membrane resonators and probe their response using laser interferometry (Fig.~\ref{fig:5}\textbf{a}). The membranes are suspended over circular cavities in SiO$_2$/Si substrates (Fig.~\ref{fig:5}\textbf{a}), with a representative membrane having a thickness of approximately $20$~nm and a diameter of $d=4$~$\mu$m. At 2 K, we observe a well-defined fundamental membrane mode mechanical resonance at $f_0=36.165$ MHz, with a quality factor of $Q \approx 1980$ (Fig.~\ref{fig:5}\textbf{c}). Since the resonance frequency of a tension-dominated membrane scales with the square root of its effective in-plane tension, it provides a sensitive probe of strain variations associated with the reconfiguration of magnetic domains~\cite{Xiao25}.

We first examine the magnetic-field dependence of the resonance frequency $f_0(H)$ after cooling the membrane to 10 K and allowing 40 hours for magnetic relaxation at zero field (Fig.~\ref{fig:5}~\textbf{d}). The resonance frequency decreases approximately symmetrically with increasing magnetic field of either polarity, indicating a reduction in the effective membrane strain $\epsilon$. This behaviour is consistent with a magnetostrictive response originating from changes in the magnetic configuration~\cite{CGT_Makars, 221_2005,Houmes2023}. Notably, a change in the slope of $f_0(H)$ (indicated by the black dashed lines) is observed around $\pm150-250$ mT, close to the field range where prominent magnetic cluster reorganisation is observed in MFM measurements for substrate-supported samples (Fig.~\ref{fig:4_MFMvsH}). Although no pronounced hysteresis is observed at this stage, the field-dependent response demonstrates that the mechanical properties of NFT are coupled to its magnetic state. Another feature is that no saturation is observed, which is consistent with other reported configurations of NFT, exhibiting the absence of the magnetisation saturation~\cite{bai2019intrinsic, qi_abnormal_2022}. 

\begin{figure*}[ht!]
\begin{center}
\includegraphics[width=2\columnwidth]{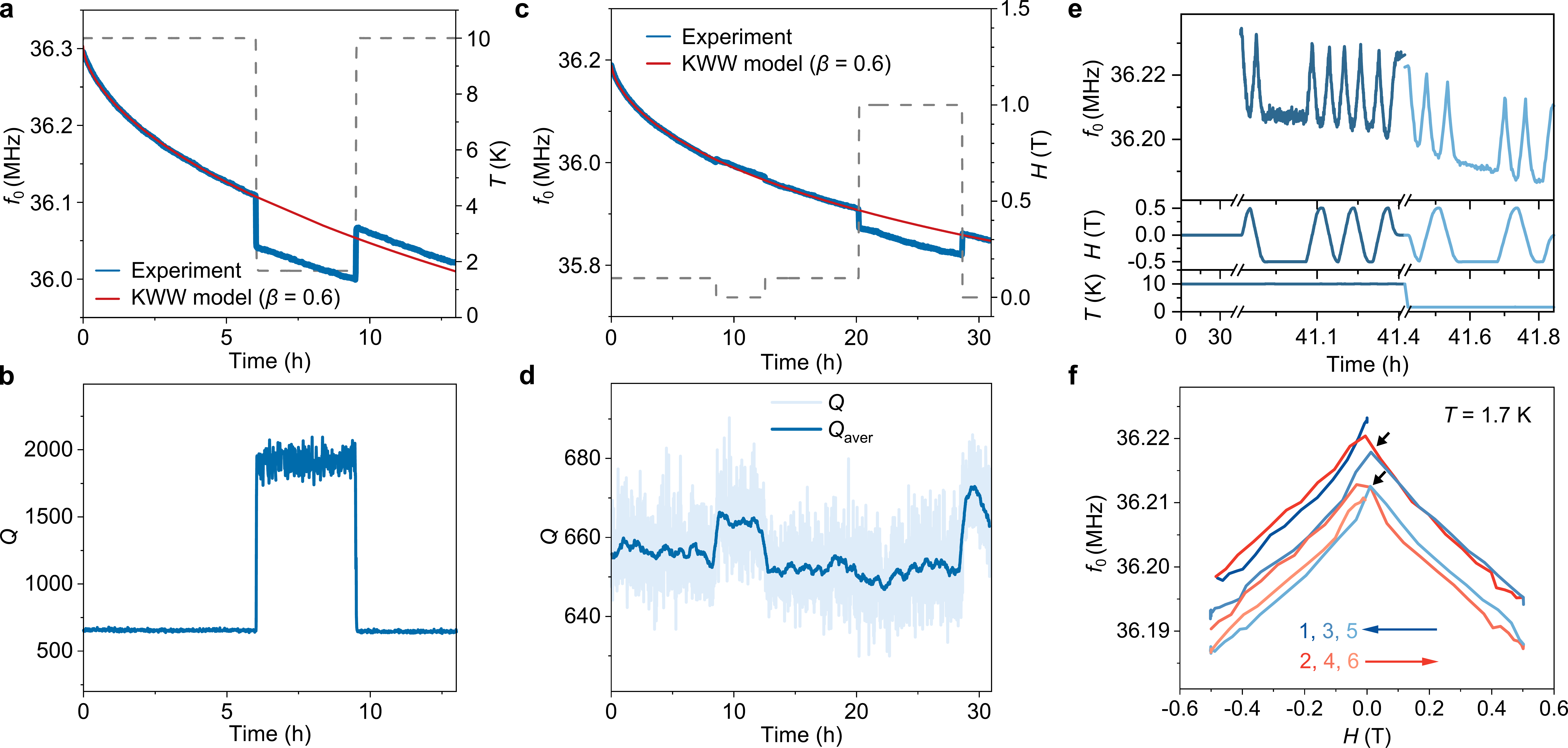}
\caption{\textbf{Mechanical fingerprints of spin-glass relaxation, ageing and history dependence.} \textbf{a-b} - Resonance frequency and quality factor during relaxation under a temporary temperature perturbation. \textbf{c-d} – Corresponding measurements during relaxation under temporary magnetic field perturbations. Red curves in \textbf{a} and \textbf{c} show fits to the Kohlrausch–Williams–Watts (KWW) stretched exponential model of equation~(\ref{fitequation}) with $\beta=0.60\pm0.01$, $\tau_{\rm ms}=72.3\pm2.8$ h in \textbf{a}, and $\beta=0.60\pm0.02$, $\tau_{\rm ms}=60.6\pm3.0$ h in \textbf{c}. Grey dashed lines indicate the temperature and magnetic field protocols, respectively. The darker blue curve in \textbf{d} represents the averaged quality factor. \textbf{e} – History of resonance frequency, magnetic field, and temperature preceding and during the measurements shown in \textbf{f}. \textbf{f} – Resonance frequency during six consecutive magnetic field sweeps at $T=1.7$ K, revealing history-dependent behaviour. Numbers indicate the sweep sequence and corresponding curve colours, while black arrows highlight regions with distinct changes in the frequency–field dependence.}
\label{fig:5_2}
\end{center}
\end{figure*}

Following that, we investigate the temperature dependence of $f_0(T)$ during consecutive heating and cooling cycles, including an intermediate waiting period at 15 K (Fig.~\ref{fig:5}~\textbf{e}). During the initial heating, $f_0(T)$ first increases at low temperatures, consistent with a negative thermal expansion contribution associated with magnetic ordering~\cite{CGT_Makars,220_2021}, before decreasing at higher temperatures due to the conventional phononic contribution to thermal expansion~\cite{112_2020} (see Supplementary Note 11). However, upon cooling, the resonance frequency follows a lower branch below approximately 50 K, coinciding with the onset of the spin glass state. This difference suggests an additional time-dependent strain contribution associated with the relaxation of magnetic clusters, similar to results in Fig.~\ref{fig:3_MFMvst}. To further investigate this effect, we introduce a waiting period at 15 K, allowing the magnetic configuration to evolve before continuing the temperature cycle. Upon subsequent reheating, the resonance frequency does not retrace the previous trajectory, indicating a memory-like response. Importantly, the frequency branches converge again within the noise level of the measurement above $T\sim50$~K, i.e. in the paramagnetic state. These observations suggest that the slowly evolving magnetic configuration below $T_g$ contributes to the membrane strain through magnetoelastic coupling, allowing the thermal history of the spin-glass state to be reflected in its mechanical response.

\subsection*{Spin-glass dynamics in nanomechanical membrane resonators}

Having established the coupling between magnetic configuration and mechanical resonance, we further investigate whether the slow relaxation of spin clusters, similar to that observed by MFM, can be quantitatively resolved through nanomechanical measurements. We first track $f_0$ over time at a constant temperature of $10$ K Fig.~\ref{fig:5_2}~\textbf{a}). The frequency exhibits a pronounced gradual decrease over several hours, indicating a continuous relaxation of the membrane tension. This behaviour can be understood in terms of magnetostriction. For a tension-dominated membrane, $f_0\propto\sqrt{\epsilon_0+\epsilon_{\rm ms}}$, where $\epsilon_0$ is the initial tensile strain and $\epsilon_{\rm ms}$ is the effective magnetostrictive contribution, typically approximated as $\epsilon_{\rm ms}\simeq\lambda_{\rm ms}M^2$ with a phenomenological magnetostriction coefficient $\lambda_{\rm ms}$~\cite{112_2020, Houmes2023}. However, in cluster spin glass materials, exchange magnetostriction is governed by local spin correlations, which do not necessarily scale with the net magnetisation. Consequently, the gradual reorganisation of magnetic clusters can modify membrane tension through exchange striction, even without a corresponding change in macroscopic $M^2$ magnetisation. 

The heterogeneous energy landscape of a cluster spin glass can give rise to a broad distribution of relaxation times, resulting in non-exponential relaxation of the associated magnetoelastic strain~\cite{Tayama2021}. It is described by the Kohlrausch-Williams-Watts (KWW) stretched exponential model~\cite{Phillips1996,Tayama2021,Matsukawa2004,Taniguchi2017}, which assumes 
\begin{equation}
\epsilon_{\rm ms}(t)= \epsilon_{\rm r}+\Delta\epsilon\;e^{-\left(\frac{t}{\tau_{\rm ms}}\right)^\beta},
\end{equation}
where $\epsilon_{\rm r}$ is the asymptotic remanent strain, $\Delta\epsilon$ is the relaxing component, $\tau_{\rm ms}$ is the characteristic relaxation timescale, and $\beta$ is the stretching exponent. Thus, the resonance frequency is
\begin{equation}\label{fitequation}
f_0(t)=\sqrt{f^2_r+\Delta f^2\;e^{-\left(\frac{t}{\tau_{\rm ms}}\right)^\beta}} ,
\end{equation}
where $f_{\rm r}$ is the frequency contribution associated with non-relaxing strain $\epsilon_0+\epsilon_{\rm r}$, and $\Delta f^2=\left(\frac{2.4048}{\pi d}\right)^2\frac{E}{\rho\left(1-\nu\right)}\Delta\epsilon$ is the relaxing component with $E$ the Young's modulus, $\rho$ the mass density, and $\nu$ the Poisson's ratio of the material. 
Remarkably, the experimental response is well described by this model with $\beta\approx0.6$. Unlike a single-exponential relaxation, which assumes one characteristic relaxation time, a stretched exponential decay with $\beta < 1$ is characteristic of a broad distribution of relaxation times~\cite{Phillips1996}. Such behaviour is consistent with the heterogeneous dynamics of a cluster spin glass~\cite{Tayama2021}, where different regions can relax through distinct metastable configurations, as observed in Fig.~\ref{fig:3_MFMvst}~\textbf{j,k}. The observed slow nanomechanical relaxation therefore provides a macroscopic manifestation of the collective spin-cluster dynamics, previously identified by MFM, in a NEMS device.

Interestingly, the gradual frequency relaxation $f_0(t)$ at constant temperature is not accompanied by a significant change in the mechanical quality factor $Q$ (Fig.~\ref{fig:5_2}~\textbf{b}). We briefly reduce the temperature from 10 K to 2 K, observing an immediate decrease in $f_0$ and a substantial increase in $Q$. During this period, the membrane tension continues to relax, while the quality factor remains approximately stable at its new value. Upon returning to 10 K, both quantities exhibit a comparable abrupt change, but the resonance frequency does not fully recover its previous trajectory. This behaviour indicates that temperature changes within the spin glass state can modify the subsequent strain relaxation pathway, consistent with the history-dependent evolution of magnetic clusters. 

We further investigate the effect of magnetic-field perturbations on the relaxation dynamics (Fig.~\ref{fig:5_2}~\textbf{c},\textbf{d}). Starting at 10 K and $H = 0.1$ T, the resonance frequency again exhibits a gradual decrease that is well described by the KWW model with $\beta \approx 0.6$ (Fig.~\ref{fig:5_2}~\textbf{c}). The similarity of the stretching exponent to that obtained during the temperature-dependent experiment suggests that both measurements probe a comparable non-single-exponential relaxation response. Briefly reducing the magnetic field to zero and returning to $0.1$ T does not substantially interrupt the gradual frequency relaxation, although a small positive change in $f_0$ and $Q$ is observed. In contrast, increasing the field to $1$ T, well beyond the characteristic magnetic-reversal field, produces an additional negative frequency shift relative to the extrapolated KWW relaxation. Restoring the field back to $0$ T recovers the previous $f_0(t)$ trajectory but causes a pronounced transient change in $Q$, which now exhibits a substantially faster decay rate than at any range of points previously. Similar nanomechanical relaxation of $f_0(t)$ is also observed in suspended graphene/NFT heterostructures (Supplementary Note 12). This suggests that sufficiently strong magnetic fields can perturb the evolving magnetic configuration, modifying the magnetoelastic strain and the corresponding mechanical response. Together with the MFM observations, these results support a picture in which the gradual reorganisation of magnetic clusters (Fig.~\ref{fig:3_MFMvst}) governs the slow relaxation of membrane tension, while external magnetic fields can modify the underlying magnetic configuration (Fig.~\ref{fig:4_MFMvsH}).

Finally, we investigate the combined influence of temperature and magnetic field history on the mechanical response (Fig.~\ref{fig:5_2}~\textbf{e},\textbf{f}). After resetting the magnetic state by heating above $T_{\rm g}$, we cool the membrane to $10$ K and allow approximately $30$ hours for relaxation. Subsequent magnetic-field sweeps between $\pm0.5$ T produce a largely reversible frequency response with only a minor gradual decrease in the resonance frequency (Fig.~\ref{fig:5_2}~\textbf{e}). We then reduce the temperature to $1.$7 K and repeat the magnetic-field sweeps without an extended waiting period. In contrast to the response at $10$ K, the successive field cycles now exhibit a downward shift of the resonance frequency, accompanied by a weak hysteresis (Fig.~\ref{fig:5_2}~\textbf{f}). Importantly, additional changes in the frequency trajectory occur around $0$ T (black arrows in Fig.~\ref{fig:5_2}~\textbf{f}), causing the system to follow progressively lower frequency branches each time the field polarity is reversed. We note that this behaviour is not observed when the system is allowed to relax for extended periods of time, e.g. immediately following $40$ hours of relaxation at $10$ K and $0$ T in Fig.~\ref{fig:5}~\textbf{d}. This indicates that the mechanical response of NFT depends not only on the applied magnetic field but also on the history of the previous magnetic state. The observed evolution suggests a combination of ongoing thermally-induced relaxation and field-induced reconfiguration of magnetic clusters, which together modify the contribution to membrane tension, effectively projecting time-, temperature-, field- and history-dependent cluster spin glass properties to the nanomechanical resonator behaviour.

\section*{Conclusion}
In summary, we have unveiled the slow spatial reorganisation of magnetic clusters in the van der Waals spin glass NbFeTe$_2$ and established NEMS as a platform for probing their complex, non-equilibrium dynamics. By integrating thin NbFeTe$_2$ flakes into suspended membrane resonators, we demonstrated that the history-dependent relaxation of magnetic clusters produces measurable shifts in mechanical resonance frequency, consistent with the evolution of internal strain mediated by magnetoelastic coupling. This mechanical readout provides a complementary probe of spin-glass physics, revealing how collective magnetic reconfigurations manifest through the elastic response of a low-dimensional material. Beyond conventional magnetic probes, the ability to resolve slow relaxation and memory-dependent behaviour through nanomechanical resonance offers opportunities to investigate the interplay between magnetic frustration, metastability and elasticity, and to explore how complex magnetic energy landscapes influence mechanical properties. More broadly, our findings establish a foundation for hybrid NEMS architectures that harness the intrinsic memory and history dependence of glassy magnetism, opening pathways towards adaptive resonators and sensors, synaptic response, and low-power unconventional information processing.

\section*{Methods}

\textbf{Crystal growth.}
NFT single crystals are grown through chemical vapour transport (CVT)~\cite{stepanova_bulk_2024, wu_tailoring_2024}. First, a polycrystalline precursor was synthesised at 750 $^\circ$C for 2 days. The resulting powder was then used as the feedstock for CVT growth under an 800 - 700 $^\circ$C temperature gradient for 14 days.
The characterisation of the grown crystals, such as Raman spectroscopy (Fig.~S1), X-ray diffraction (Fig.~S2), and Energy-dispersive X-ray (EDX) (Fig.~S3), is presented in the SI.
EDX analysis of the synthesised flakes revealed a stoichiometric composition of Nb$_{0.91}$Fe$_{1.13}$Te$_2$ (Fig.~S3). The peak positions and their relative intensities closely resemble those of the LT-phase observed in Ref.~\cite{wu_tailoring_2024}.

\textbf{Vibrating Sample Magnetometry.}
Vibrating Sample Magnetometry (VSM) was performed using a Quantum Design MPMS 3. We used \ce{SiO2}/Si with double-sided Kapton adhesive tape mounted inside the plastic straw as a substrate holding the bulk crystal of NFT. All presented curves were captured with the bare substrate without the crystal characterised first to ensure a negligible contribution to the total magnetic moment. The field-cooling sequence included ramping the magnetic field to 100 Oe at room temperature, cooling the sample down to 10 K with 10~K/min, then to 2~K with 1.6 K/min. After reaching 2~K, the temperature was raised back to RT, and the total magnetic moment was captured with 150-point resolution. Zero-field cooling was performed with the same temperature ramping speeds. To capture $M(H)$ field was ramped with the sequence {$H_\mathrm{max}$ $\rightarrow$ $-H_\mathrm{max}$ $\rightarrow$ $H_\mathrm{max}$. The field ramp rate was 200 G/s. AC measurements were performed with an oscillating AC driving magnetic field of 5~Oe in a small static field of 0.5~Oe.

\textbf{Atomistic simulations.} The spin dynamics of \ce{NbFeTe2} magnets was investigated using atomistic simulation methods\cite{evans_atomistic_2014,Kartsev20,Wahab2021,Zhou25,Xiao25,Haldar25}. For autocorrelation calculations, the system considers a \(\mathrm{50\times50\times15\:nm^3}\) computational cell with an orthorhombic lattice structure. Periodic boundary conditions were applied in the x- and y-dimensions to minimise finite-size effects. Exchange interactions were calculated via first-principles techniques up to a cutoff range of 5 \AA \: that includes the first three nearest-neighbour interactions. Simulations were performed using the Landau-Lifshitz-Gilbert (LLG) equation, integrated numerically with the Heun scheme. See Supplementary Note 4 for additional details. 

\textbf{Sample fabrication for MCD and MFM.}
Thin \ce{NbFeTe2} samples for MCD and MFM measurements were mechanically exfoliated on a \ce{SiO2}/Si substrate inside an Ar glovebox with a concentration of water and oxygen of less than 0.5 ppm. By mechanical exfoliation, it was possible to obtain samples with a minimum thickness of $\sim$10~nm.

\textbf{Magnetic Circular Dichroism.}
In the MCD experiments, a continuous wave (CW) laser with a wavelength of 532 nm (MGL-S-532/200~mW) was used. To facilitate the switching of incident light between left- and right-handed circular polarisation, a polariser (Thorlabs GTH10M) and a photoelastic modulator (PEM, Hinds Instruments I/FS50) were positioned at a 45$^\circ$ angle relative to each other. The PEM was configured to operate in quarter-wave retardation mode. The laser beam was directed by a nonpolarizing beamsplitter (Lambda Research Optics BNPB-25.4B-45R-800) into a closed-cycle liquid helium cryostat (Quantum Design PPMS) equipped with a 9~T magnet and in attocube attoDRY 2100. It was then focused using an objective lens (attocube LT-APO VISNIR 0.82). The incident laser power was set at 6 $\upmu$W. The laser reflected from the sample was collected by the same lens, passed through the beamsplitter, and subsequently detected by a photodiode (Thorlabs DET36A/M). The signal from the photodiode was measured using a lock-in amplifier (Stanford Research Systems SR860) tuned to the first harmonic (1f) of the PEM.

\textbf{Magnetic force microscopy.}
Atomic force microscopy (AFM) and magnetic force microscopy (MFM) were performed using the attoAFM I probe in the Attocube AttoDRY 2100 closed-cycle cryostat operating at a base temperature of 1.7~K, equipped with and featuring a superconducting magnet capable of reaching up to 9~T. Silicon cantilevers with magnetic CoCr coating were used, with a spring constant $k$ of $2-4$~N/m and tip radius $30-50$~nm (PPP-MFMR and SSS-MFMR from Nanosensors). Before measurements, the cantilevers were magnetised at room temperature using a neodymium magnet in the out-of-plane direction. The magnetic signal from the sample was detected in non-contact mode in the phase channel.

\textbf{MFM image analysis.}
For multiscale structural complexity analysis, MFM images were converted to grayscale. This was done by applying a normalised colour scale for each image. For local complexity calculations in Fig.~\ref{fig:3_MFMvst}, the general scale range was set between -0.4 and 0.4 degrees, whereas for the entire image, it was set between -0.7 and 0.7 degrees. Subsequently, the complexity calculation procedure and the code from Bagrov, A. et al~\cite{bagrov2020multiscale} were applied (see Supplementary Note 8).

MFM phase range was extracted based on the minimum and maximum values of the MFM phase with a tail cutoff, fixing the range not taking into account the lowest and largest 0.1~$\%$ values to minimise the influence of spikes or rare points in the data analysis (see Supplementary Note~9).

Autocorrelation length $\xi$ was calculated using:
\begin{equation}
C(\mathbf{r})=\left\langle \delta\phi(\mathbf{r}'+\mathbf{r})\,\delta\phi(\mathbf{r}')\right\rangle_{\mathbf{r}'},
\qquad
\frac{\overline{C}(\xi)}{\overline{C}(0^{+})}=\frac{1}{e},
\end{equation}

where $\delta\phi=\phi-\langle\phi\rangle$ is the MFM phase with its mean removed, $\overline{C}(r)$ is the azimuthal average of $C(\mathbf{r})$, and $\overline{C}(0^{+})$ is its extrapolation to $r\to0$, which excludes the uncorrelated noise contribution at $r=0$.

\textbf{Membrane sample fabrication.}
\ce{SiO2}/Si substrates were patterned by electron-beam lithography (EBL) and subsequently etched by deep reactive ion etching (Deep RIE) to produce circular cavities with diameters of 1–4~$\upmu$m and a depth of 285~nm. Flakes were then mechanically exfoliated and transferred onto the substrates using PDMS stamps inside an Ar glovebox.

\textbf{Laser interferometry measurements.}
A blue diode laser ($\lambda_{\rm{blue}}$ = 450 nm), power-modulated by a R$\&$S$^\text\textregistered$ ZNL3 vector network analyser (VNA), was used to excite the membrane and drive it into motion. The displacement was detected by focusing a red He-Ne laser ($\lambda_{\rm{red}}$ = 633 nm) onto the cavity beneath the membrane. The reflected light, modulated by the membrane motion, was recorded by a photodiode and analysed by the phase-sensitive VNA. The measurement conditions were verified to ensure that the laser-induced heating did not significantly affect the results.

\subsection*{Data availability}
All data supporting the findings of this article and its Supplementary Information will be made available upon request to the authors.

\section*{Acknowledgements}
This research / project is supported by the Ministry of Education, Singapore, under its funding for the Research Centre of Excellence award to the Institute for Functional Intelligent Materials, Project No. EDUNC-33-18-279-V12-IFIM. This research is supported by the Ministry of Education, Singapore, under its Academic Research Fund Tier 2 (MOE-T2EP50122-0012). This material is based upon work supported by the Air Force Office of Scientific Research and the Office of Naval Research Global under award number FA8655-21-1-7026.
P.C. would like to acknowledge the China Scholarship Council (CSC) Grant No. 202208060246 for sponsoring his work. E.J.G.S. acknowledges computational resources through CIRRUS Tier-2 HPC Service (ec131, e980) at EPCC (http://www.cirrus.ac.uk) funded by the University of Edinburgh and EPSRC (EP/P020267/1); and ARCHER2 UK National Supercomputing Service via the UKCP consortium (Project e89) funded by EPSRC grant ref EP/X035891/1. E.J.G.S. acknowledges the EPSRC Open Fellowship (EP/T021578/1) and the Donostia International Physics Center for funding support. E.J.G.S. also acknowledges the Basque Government’s BERC and the Severo Ochoa Centre of Excellence Programmes (CEX2024-001491-S, funded by MICIU/AEI/10.13039/501100011033). We gratefully acknowledge the computing resources provided on Swing (Project MagSim), a high-performance computing cluster operated by the Laboratory Computing Resource Center at Argonne National Laboratory.
Z. Liu acknowledges support from the Singapore Ministry of Education Tier 3 and Tier 2 Programmatic Fund (award no. MOEMOET32023-0003, and MOE-T2EP50223-0008). R. Duan acknowledges support from the Agency for Science, Technology and Research (A*STAR) under the MTC Young Individual Research Grants (YIRG) (M24N8c0097). G.E. acknowledges support from the Ministry of Education (MOE), Singapore, under AcRF Tier 2 (T2EP50223-0023) and Tier 1 (A-8001995-00-00). 
B.\"O. acknowledges support by the National Research Foundation, Prime Minister's Office, Singapore, under its Competitive Research Programme (CRP award number NRF‐CRP22‐2019‐008 and Frontier CRP award number NRF‐F-CRP-2024-0010).

\hspace{0.5cm}
\section*{Author Contributions}
S.Y.G. and M.\v{S}. conceived the project. S.Y.G., M.\v{S}., E.J.G.S., M.K., and K.S.N. supervised the research. R.D. and L.Z. provided grown bulk NFT crystals with Raman, XRD and EDX characterisation. Y.W.H. and G.E. conducted MCD measurements. A.K.G. and B.\"O. supported with VSM measurements of bulk NFT. Z.Y., P.C., S.R., and E.J.G.S. conducted atomistic simulations. S.Y.G. performed MFM measurements. S.Y.G. analyzed VSM, MCD and MFM data. M.\v{S}. fabricated and prepared samples for VSM, MCD, MFM, and nanomechanical measurements. T.M.S., D.A.L., J.M.W., and M.\v{S}. performed nanomechanical measurements. T.M.S., M.\v{S}. conducted data analysis of nanomechanics data. M.Kr. wrote software for nanomechanical probing. The paper was jointly written by all authors, with the main contribution from S.Y.G. and M.\v{S}. All authors discussed the results and commented on the paper.

\section*{Competing interests}

The authors declare no conflict of interest.

\bibliographystyle{naturemag}

\begin{thebibliography}{10}
\expandafter\ifx\csname url\endcsname\relax
  \def\url#1{\texttt{#1}}\fi
\expandafter\ifx\csname urlprefix\endcsname\relax\def\urlprefix{URL }\fi
\providecommand{\bibinfo}[2]{#2}
\providecommand{\eprint}[2][]{\url{#2}}

\bibitem{edwards_theory_1975}
\bibinfo{author}{Edwards, S.~F.} \& \bibinfo{author}{Anderson, P.~W.}
\newblock \bibinfo{title}{Theory of spin glasses}.
\newblock \emph{\bibinfo{journal}{Journal of Physics F: Metal Physics}} \textbf{\bibinfo{volume}{5}}, \bibinfo{pages}{965} (\bibinfo{year}{1975}).

\bibitem{edwards_theory_1976}
\bibinfo{author}{Edwards, S.~F.} \& \bibinfo{author}{Anderson, P.~W.}
\newblock \bibinfo{title}{Theory of spin glasses. {II}}.
\newblock \emph{\bibinfo{journal}{Journal of Physics F: Metal Physics}} \textbf{\bibinfo{volume}{6}}, \bibinfo{pages}{1927} (\bibinfo{year}{1976}).

\bibitem{binder1986spin}
\bibinfo{author}{Binder, K.} \& \bibinfo{author}{Young, A.~P.}
\newblock \bibinfo{title}{Spin glasses: Experimental facts, theoretical concepts, and open questions}.
\newblock \emph{\bibinfo{journal}{Reviews of Modern Physics}} \textbf{\bibinfo{volume}{58}}, \bibinfo{pages}{801} (\bibinfo{year}{1986}).

\bibitem{altieri_introduction_2024}
\bibinfo{author}{Altieri, A.} \& \bibinfo{author}{Baity-Jesi, M.}
\newblock \bibinfo{title}{{An introduction to the theory of spin glasses}}.
\newblock In \emph{\bibinfo{booktitle}{{Encyclopedia of Condensed Matter Physics}}}, \bibinfo{pages}{361--370} (\bibinfo{publisher}{{Elsevier}}, \bibinfo{year}{2024}).

\bibitem{hertz2018introduction}
\bibinfo{author}{Hertz, J.~A.}, \bibinfo{author}{Krogh, A.} \& \bibinfo{author}{Palmer, R.~G.}
\newblock \emph{\bibinfo{title}{Introduction to the theory of neural computation}} (\bibinfo{publisher}{Crc Press}, \bibinfo{year}{2018}).

\bibitem{burch_magnetism_2018}
\bibinfo{author}{Burch, K.~S.}, \bibinfo{author}{Mandrus, D.} \& \bibinfo{author}{Park, J.~G.}
\newblock \bibinfo{title}{Magnetism in two-dimensional van der {Waals} materials}.
\newblock \emph{\bibinfo{journal}{Nature}} \textbf{\bibinfo{volume}{563}}, \bibinfo{pages}{47--52} (\bibinfo{year}{2018}).

\bibitem{Gibertini2019}
\bibinfo{author}{Gibertini, M.}, \bibinfo{author}{Koperski, M.}, \bibinfo{author}{Morpurgo, A.~F.} \& \bibinfo{author}{Novoselov, K.~S.}
\newblock \bibinfo{title}{Magnetic {2D} materials and heterostructures}.
\newblock \emph{\bibinfo{journal}{Nature Nanotechnology}} \textbf{\bibinfo{volume}{14}}, \bibinfo{pages}{408--419} (\bibinfo{year}{2019}).

\bibitem{Gong2019}
\bibinfo{author}{Gong, C.} \& \bibinfo{author}{Zhang, X.}
\newblock \bibinfo{title}{Two-dimensional magnetic crystals and emergent heterostructure devices}.
\newblock \emph{\bibinfo{journal}{Science}} \textbf{\bibinfo{volume}{363}}, \bibinfo{pages}{eaav4450} (\bibinfo{year}{2019}).

\bibitem{Mak2019}
\bibinfo{author}{Mak, K.~F.}, \bibinfo{author}{Shan, J.} \& \bibinfo{author}{Ralph, D.~C.}
\newblock \bibinfo{title}{Probing and controlling magnetic states in {2D} layered magnetic materials}.
\newblock \emph{\bibinfo{journal}{Nature Reviews Physics}} \textbf{\bibinfo{volume}{1}}, \bibinfo{pages}{646--661} (\bibinfo{year}{2019}).

\bibitem{huang_emergent_2020}
\bibinfo{author}{Huang, B.} \emph{et~al.}
\newblock \bibinfo{title}{Emergent phenomena and proximity effects in two-dimensional magnets and heterostructures}.
\newblock \emph{\bibinfo{journal}{Nature Materials}} \textbf{\bibinfo{volume}{19}}, \bibinfo{pages}{1276--1289} (\bibinfo{year}{2020}).

\bibitem{kurebayashi_magnetism_2022}
\bibinfo{author}{Kurebayashi, H.}, \bibinfo{author}{Garcia, J.~H.}, \bibinfo{author}{Khan, S.}, \bibinfo{author}{Sinova, J.} \& \bibinfo{author}{Roche, S.}
\newblock \bibinfo{title}{Magnetism, symmetry and spin transport in van der {Waals} layered systems}.
\newblock \emph{\bibinfo{journal}{Nature Reviews Physics}} \textbf{\bibinfo{volume}{4}}, \bibinfo{pages}{150--166} (\bibinfo{year}{2022}).

\bibitem{Genome22}
\bibinfo{author}{Wang, Q.~H.} \emph{et~al.}
\newblock \bibinfo{title}{{The Magnetic Genome of Two-Dimensional van der {W}aals Materials}}.
\newblock \emph{\bibinfo{journal}{ACS Nano}} \textbf{\bibinfo{volume}{16}}, \bibinfo{pages}{6960--7079} (\bibinfo{year}{2022}).

\bibitem{TANG20231}
\bibinfo{author}{Tang, C.} \emph{et~al.}
\newblock \bibinfo{title}{{Spin dynamics in van der {W}aals magnetic systems}}.
\newblock \emph{\bibinfo{journal}{Physics Reports}} \textbf{\bibinfo{volume}{1032}}, \bibinfo{pages}{1--36} (\bibinfo{year}{2023}).

\bibitem{112_2020}
\bibinfo{author}{Šiškins, M.} \emph{et~al.}
\newblock \bibinfo{title}{Magnetic and electronic phase transitions probed by nanomechanical resonators}.
\newblock \emph{\bibinfo{journal}{Nature Communications}} \textbf{\bibinfo{volume}{11}}, \bibinfo{pages}{2698} (\bibinfo{year}{2020}).

\bibitem{Jiang2020}
\bibinfo{author}{Jiang, S.}, \bibinfo{author}{Xie, H.}, \bibinfo{author}{Shan, J.} \& \bibinfo{author}{Mak, K.~F.}
\newblock \bibinfo{title}{Exchange magnetostriction in two-dimensional antiferromagnets}.
\newblock \emph{\bibinfo{journal}{Nature Materials}} \textbf{\bibinfo{volume}{19}}, \bibinfo{pages}{1295–1299} (\bibinfo{year}{2020}).

\bibitem{CGT_Makars}
\bibinfo{author}{Šiškins, M.} \emph{et~al.}
\newblock \bibinfo{title}{Nanomechanical probing and strain tuning of the {C}urie temperature in suspended {Cr$_2$Ge$_2$Te$_6$}-based heterostructures}.
\newblock \emph{\bibinfo{journal}{npj 2D Materials and Applications}} \textbf{\bibinfo{volume}{6}}, \bibinfo{pages}{41} (\bibinfo{year}{2022}).

\bibitem{pal2024realization}
\bibinfo{author}{Pal, B.} \emph{et~al.}
\newblock \bibinfo{title}{Realization of a spin glass in a two-dimensional van der {W}aals material}.
\newblock \emph{\bibinfo{journal}{Science}} \textbf{\bibinfo{volume}{392}}, \bibinfo{pages}{624–628} (\bibinfo{year}{2026}).

\bibitem{khan2024spin}
\bibinfo{author}{Khan, S.} \emph{et~al.}
\newblock \bibinfo{title}{Spin-glass states generated in a van der {W}aals magnet by alkali-ion intercalation}.
\newblock \emph{\bibinfo{journal}{Advanced Materials}} \textbf{\bibinfo{volume}{36}}, \bibinfo{pages}{2400270} (\bibinfo{year}{2024}).

\bibitem{takano2003magnetic}
\bibinfo{author}{Takano, Y.}, \bibinfo{author}{Arai, A.}, \bibinfo{author}{Takahashi, Y.}, \bibinfo{author}{Takase, K.} \& \bibinfo{author}{Sekizawa, K.}
\newblock \bibinfo{title}{Magnetic properties and specific heat of new spin glass {Mn$_{0.5}$Fe$_{0.5}$PS$_{3}$}}.
\newblock \emph{\bibinfo{journal}{Journal of Applied Physics}} \textbf{\bibinfo{volume}{93}}, \bibinfo{pages}{8197--8199} (\bibinfo{year}{2003}).

\bibitem{masubuchi2008phase}
\bibinfo{author}{Masubuchi, T.} \emph{et~al.}
\newblock \bibinfo{title}{Phase diagram, magnetic properties and specific heat of {Mn$_{1-x}$Fe$_x$PS$_3$}}.
\newblock \emph{\bibinfo{journal}{Journal of Alloys and Compounds}} \textbf{\bibinfo{volume}{460}}, \bibinfo{pages}{668--674} (\bibinfo{year}{2008}).

\bibitem{goossens2013magnetic}
\bibinfo{author}{Goossens, D.}, \bibinfo{author}{Brazier-Hollins, S.}, \bibinfo{author}{James, D.}, \bibinfo{author}{Hutchison, W.} \& \bibinfo{author}{Hester, J.}
\newblock \bibinfo{title}{Magnetic structure and glassiness in {Fe$_{0.5}$Ni$_{0.5}$PS$_3$}}.
\newblock \emph{\bibinfo{journal}{Journal of Magnetism and Magnetic Materials}} \textbf{\bibinfo{volume}{334}}, \bibinfo{pages}{82--86} (\bibinfo{year}{2013}).

\bibitem{graham2020local}
\bibinfo{author}{Graham, J.} \emph{et~al.}
\newblock \bibinfo{title}{Local nuclear and magnetic order in the two-dimensional spin glass {Mn$_{0.5}$Fe$_{0.5}$PS$_3$}}.
\newblock \emph{\bibinfo{journal}{Physical Review Materials}} \textbf{\bibinfo{volume}{4}}, \bibinfo{pages}{084401} (\bibinfo{year}{2020}).

\bibitem{mcguire2018spin}
\bibinfo{author}{McGuire, M.~A.} \& \bibinfo{author}{Sales, B.~C.}
\newblock \bibinfo{title}{Spin-glass behavior and vacancy order in van der {W}aals layered $\beta$-{MoCl}$_4$}.
\newblock \emph{\bibinfo{journal}{Physical Review Materials}} \textbf{\bibinfo{volume}{2}}, \bibinfo{pages}{074007} (\bibinfo{year}{2018}).

\bibitem{wang2023spin}
\bibinfo{author}{Wang, X.} \emph{et~al.}
\newblock \bibinfo{title}{Spin glass behavior in amorphous {Cr$_2$Ge$_2$Te$_6$} phase-change alloy}.
\newblock \emph{\bibinfo{journal}{Advanced Science}} \textbf{\bibinfo{volume}{10}}, \bibinfo{pages}{2302444} (\bibinfo{year}{2023}).

\bibitem{torres2024glassy}
\bibinfo{author}{Torres, K.} \emph{et~al.}
\newblock \bibinfo{title}{Glassy relaxation dynamics in the two-dimensional heavy fermion antiferromagnet {CeSiI}}.
\newblock \emph{\bibinfo{journal}{Nano Letters}} \textbf{\bibinfo{volume}{25}}, \bibinfo{pages}{6848–6854} (\bibinfo{year}{2025}).

\bibitem{zhang_magnetic_1997}
\bibinfo{author}{Zhang, J.~H.}, \bibinfo{author}{Chen, F.}, \bibinfo{author}{Li, J.} \& \bibinfo{author}{O’Connor, C.~J.}
\newblock \bibinfo{title}{Magnetic property of layered compound {NbFeTe$_2$}}.
\newblock \emph{\bibinfo{journal}{Journal of Applied Physics}} \textbf{\bibinfo{volume}{81}}, \bibinfo{pages}{5283--5285} (\bibinfo{year}{1997}).

\bibitem{stepanova_bulk_2024}
\bibinfo{author}{Stepanova, A.~V.} \emph{et~al.}
\newblock \bibinfo{title}{Bulk ferromagnetism in cleavable van der {Waals} telluride {NbFeTe$_2$}}.
\newblock \emph{\bibinfo{journal}{Chemical Communications}} \textbf{\bibinfo{volume}{60}}, \bibinfo{pages}{5518--5521} (\bibinfo{year}{2024}).

\bibitem{wu_tailoring_2024}
\bibinfo{author}{Wu, H.} \emph{et~al.}
\newblock \bibinfo{title}{Tailoring physical properties of crystals through synthetic temperature control: {A} case study for new polymorphic {NbFeTe}$_{2}$ phases}.
\newblock \emph{\bibinfo{journal}{Physical Review B}} \textbf{\bibinfo{volume}{109}}, \bibinfo{pages}{174427} (\bibinfo{year}{2024}).

\bibitem{Xiao25}
\bibinfo{author}{Yousuf, S. M. E.~H.} \emph{et~al.}
\newblock \bibinfo{title}{Mechanical resonant sensing of spin texture dynamics in a {2D} antiferromagnet}.
\newblock \emph{\bibinfo{journal}{Advanced Materials}} \textbf{\bibinfo{volume}{37}}, \bibinfo{pages}{2420168} (\bibinfo{year}{2025}).

\bibitem{208_2022}
\bibinfo{author}{Li, X.}, \bibinfo{author}{Sha, X.}, \bibinfo{author}{Yan, N.} \& \bibinfo{author}{Zhang, T.}
\newblock \bibinfo{title}{Mechanical detection of magnetic phase transition in suspended {CrOCl} heterostructures}.
\newblock \emph{\bibinfo{journal}{Magnetochemistry}} \textbf{\bibinfo{volume}{8}}, \bibinfo{pages}{170} (\bibinfo{year}{2022}).

\bibitem{213_2024}
\bibinfo{author}{Fei, F.} \emph{et~al.}
\newblock \bibinfo{title}{Spin-mechanical coupling in {2D} antiferromagnet {CrSBr}}.
\newblock \emph{\bibinfo{journal}{Nano Letters}} \textbf{\bibinfo{volume}{24}}, \bibinfo{pages}{10467--10474} (\bibinfo{year}{2024}).

\bibitem{gunnarsson1988dynamics}
\bibinfo{author}{Gunnarsson, K.} \emph{et~al.}
\newblock \bibinfo{title}{Dynamics of an ising spin-glass in the vicinity of the spin-glass temperature}.
\newblock \emph{\bibinfo{journal}{Physical Review Letters}} \textbf{\bibinfo{volume}{61}}, \bibinfo{pages}{754} (\bibinfo{year}{1988}).

\bibitem{mydosh1993spin}
\bibinfo{author}{Mydosh, J.~A.}
\newblock \emph{\bibinfo{title}{Spin glasses: an experimental introduction}} (\bibinfo{publisher}{{CRC} press}, \bibinfo{year}{1993}).

\bibitem{benka2022interplay}
\bibinfo{author}{Benka, G.} \emph{et~al.}
\newblock \bibinfo{title}{Interplay of itinerant magnetism and spin-glass behavior in {Fe$_x$Cr$_{1-x}$}}.
\newblock \emph{\bibinfo{journal}{Physical Review Materials}} \textbf{\bibinfo{volume}{6}}, \bibinfo{pages}{044407} (\bibinfo{year}{2022}).

\bibitem{malinowski2011spin}
\bibinfo{author}{Malinowski, A.} \emph{et~al.}
\newblock \bibinfo{title}{Spin-glass behavior in {N}i-doped {La$_{1.85}$Sr$_{0.15}$CuO$_4$}}.
\newblock \emph{\bibinfo{journal}{Physical Review B}} \textbf{\bibinfo{volume}{84}}, \bibinfo{pages}{024409} (\bibinfo{year}{2011}).

\bibitem{bag2020study}
\bibinfo{author}{Bag, P.}, \bibinfo{author}{Somesh, K.} \& \bibinfo{author}{Nath, R.}
\newblock \bibinfo{title}{A study of cluster spin-glass behaviour at the critical composition {Mn$_{0.73}$Fe$_{0.27}$NiGe}}.
\newblock \emph{\bibinfo{journal}{Journal of Magnetism and Magnetic Materials}} \textbf{\bibinfo{volume}{497}}, \bibinfo{pages}{165977} (\bibinfo{year}{2020}).

\bibitem{henkel2007ageing}
\bibinfo{author}{Henkel, M.}, \bibinfo{author}{Pleimling, M.} \& \bibinfo{author}{Sanctuary, R.}
\newblock \emph{\bibinfo{title}{Ageing and the glass transition}}, vol. \bibinfo{volume}{716} (\bibinfo{publisher}{Springer Science \& Business Media}, \bibinfo{year}{2007}).

\bibitem{mydosh2015spin}
\bibinfo{author}{Mydosh, J.~A.}
\newblock \bibinfo{title}{Spin glasses: redux: an updated experimental/materials survey}.
\newblock \emph{\bibinfo{journal}{Reports on Progress in Physics}} \textbf{\bibinfo{volume}{78}}, \bibinfo{pages}{052501} (\bibinfo{year}{2015}).

\bibitem{Kartsev20}
\bibinfo{author}{Kartsev, A.}, \bibinfo{author}{Augustin, M.}, \bibinfo{author}{Evans, R. F.~L.}, \bibinfo{author}{Novoselov, K.~S.} \& \bibinfo{author}{Santos, E. J.~G.}
\newblock \bibinfo{title}{Biquadratic exchange interactions in two-dimensional magnets}.
\newblock \emph{\bibinfo{journal}{npj Computational Materials}} \textbf{\bibinfo{volume}{6}}, \bibinfo{pages}{150} (\bibinfo{year}{2020}).

\bibitem{Wahab2021}
\bibinfo{author}{Wahab, D.~A.} \emph{et~al.}
\newblock \bibinfo{title}{{Quantum Rescaling, Domain Metastability, and Hybrid Domain-Walls in {2D} CrI$_3$ Magnets}}.
\newblock \emph{\bibinfo{journal}{Advanced Materials}} \textbf{\bibinfo{volume}{33}}, \bibinfo{pages}{2004138} (\bibinfo{year}{2021}).

\bibitem{Zhou25}
\bibinfo{author}{Wang, Y.-X.} \emph{et~al.}
\newblock \bibinfo{title}{Configurable antiferromagnetic domains and lateral exchange bias in atomically thin {CrPS$_4$}}.
\newblock \emph{\bibinfo{journal}{Nature Materials}} \textbf{\bibinfo{volume}{24}}, \bibinfo{pages}{1414–1423} (\bibinfo{year}{2025}).

\bibitem{Haldar25}
\bibinfo{author}{Haldar, S.}, \bibinfo{author}{Griepe, T.}, \bibinfo{author}{Atxitia, U.} \& \bibinfo{author}{Santos, E. J.~G.}
\newblock \bibinfo{title}{All-heat control of magnetization dynamics on van der {W}aals magnets}.
\newblock \emph{\bibinfo{journal}{Advanced Materials}} \textbf{\bibinfo{volume}{37}}, \bibinfo{pages}{2501043} (\bibinfo{year}{2025}).

\bibitem{kamber_self-induced_2020}
\bibinfo{author}{Kamber, U.} \emph{et~al.}
\newblock \bibinfo{title}{Self-induced spin glass state in elemental and crystalline neodymium}.
\newblock \emph{\bibinfo{journal}{Science}} \textbf{\bibinfo{volume}{368}}, \bibinfo{pages}{eaay6757} (\bibinfo{year}{2020}).

\bibitem{verlhac_thermally_2022}
\bibinfo{author}{Verlhac, B.} \emph{et~al.}
\newblock \bibinfo{title}{Thermally induced magnetic order from glassiness in elemental neodymium}.
\newblock \emph{\bibinfo{journal}{Nature Physics}} \textbf{\bibinfo{volume}{18}}, \bibinfo{pages}{905--911} (\bibinfo{year}{2022}).

\bibitem{kroeze_directly_2024}
\bibinfo{author}{Kroeze, R.~M.} \emph{et~al.}
\newblock \bibinfo{title}{Directly observing replica symmetry breaking in a vector quantum-optical spin glass}.
\newblock \emph{\bibinfo{journal}{Science}} \textbf{\bibinfo{volume}{389}}, \bibinfo{pages}{1122–1126} (\bibinfo{year}{2025}).

\bibitem{bagrov2020multiscale}
\bibinfo{author}{Bagrov, A.~A.}, \bibinfo{author}{Iakovlev, I.~A.}, \bibinfo{author}{Iliasov, A.~A.}, \bibinfo{author}{Katsnelson, M.~I.} \& \bibinfo{author}{Mazurenko, V.~V.}
\newblock \bibinfo{title}{Multiscale structural complexity of natural patterns}.
\newblock \emph{\bibinfo{journal}{Proceedings of the National Academy of Sciences}} \textbf{\bibinfo{volume}{117}}, \bibinfo{pages}{30241--30251} (\bibinfo{year}{2020}).

\bibitem{221_2005}
\bibinfo{author}{Masmanidis, S.~C.} \emph{et~al.}
\newblock \bibinfo{title}{Nanomechanical measurement of magnetostriction and magnetic anisotropy in {(Ga,Mn)As}}.
\newblock \emph{\bibinfo{journal}{Physical Review Letters}} \textbf{\bibinfo{volume}{95}}, \bibinfo{pages}{187206} (\bibinfo{year}{2005}).

\bibitem{Houmes2023}
\bibinfo{author}{Houmes, M. J.~A.} \emph{et~al.}
\newblock \bibinfo{title}{Magnetic order in {2D} antiferromagnets revealed by spontaneous anisotropic magnetostriction}.
\newblock \emph{\bibinfo{journal}{Nature Communications}} \textbf{\bibinfo{volume}{14}}, \bibinfo{pages}{8503} (\bibinfo{year}{2023}).

\bibitem{bai2019intrinsic}
\bibinfo{author}{Bai, W.} \emph{et~al.}
\newblock \bibinfo{title}{Intrinsic negative magnetoresistance in van der {W}aals {FeNbTe$_2$} single crystals}.
\newblock \emph{\bibinfo{journal}{Advanced Materials}} \textbf{\bibinfo{volume}{31}}, \bibinfo{pages}{1900246} (\bibinfo{year}{2019}).

\bibitem{qi_abnormal_2022}
\bibinfo{author}{Qi, B.-t.} \emph{et~al.}
\newblock \bibinfo{title}{Abnormal {Magnetoresistance} {Transport} {Properties} of van der {Waals} {Antiferromagnetic} {FeNbTe$_2$}}.
\newblock \emph{\bibinfo{journal}{Frontiers in Physics}} \textbf{\bibinfo{volume}{10}}, \bibinfo{pages}{851838} (\bibinfo{year}{2022}).

\bibitem{220_2021}
\bibinfo{author}{Kozlenko, D.~P.} \emph{et~al.}
\newblock \bibinfo{title}{Spin-induced negative thermal expansion and spin–phonon coupling in van der {W}aals material {CrBr$_3$}}.
\newblock \emph{\bibinfo{journal}{npj Quantum Materials}} \textbf{\bibinfo{volume}{6}}, \bibinfo{pages}{19} (\bibinfo{year}{2021}).

\bibitem{Tayama2021}
\bibinfo{author}{Tayama, T.}, \bibinfo{author}{Matsuda, T.~D.}, \bibinfo{author}{Higashinaka, R.} \& \bibinfo{author}{Aoki, Y.}
\newblock \bibinfo{title}{Magnetization and thermal expansion measurements on the {Ising} magnet {SmPt$_2$Si$_2$}: {p}otential coexistence of spin glass and antiferromagnetic states}.
\newblock \emph{\bibinfo{journal}{Physical Review B}} \textbf{\bibinfo{volume}{104}}, \bibinfo{pages}{174418} (\bibinfo{year}{2021}).

\bibitem{Phillips1996}
\bibinfo{author}{Phillips, J.~C.}
\newblock \bibinfo{title}{Stretched exponential relaxation in molecular and electronic glasses}.
\newblock \emph{\bibinfo{journal}{Reports on Progress in Physics}} \textbf{\bibinfo{volume}{59}}, \bibinfo{pages}{1133–1207} (\bibinfo{year}{1996}).

\bibitem{Matsukawa2004}
\bibinfo{author}{Matsukawa, M.} \emph{et~al.}
\newblock \bibinfo{title}{Stretched exponential behavior in remanent lattice striction of a {(La,Pr)$_{1.2}$Sr$_{1.8}$Mn$_2$O$_7$} bilayer manganite single crystal}.
\newblock \emph{\bibinfo{journal}{Physical Review B}} \textbf{\bibinfo{volume}{70}}, \bibinfo{pages}{132402} (\bibinfo{year}{2004}).

\bibitem{Taniguchi2017}
\bibinfo{author}{Taniguchi, H.} \emph{et~al.}
\newblock \bibinfo{title}{Dynamics of the lattice and spins in the phase-separated manganite {(Eu$_{1-x}$Gd$_x$)$_{0.6}$Sr$_{0.4}$MnO$_3$}}.
\newblock \emph{\bibinfo{journal}{Journal of Magnetism and Magnetic Materials}} \textbf{\bibinfo{volume}{432}}, \bibinfo{pages}{548--553} (\bibinfo{year}{2017}).

\bibitem{evans_atomistic_2014}
\bibinfo{author}{Evans, R. F.~L.} \emph{et~al.}
\newblock \bibinfo{title}{Atomistic spin model simulations of magnetic nanomaterials}.
\newblock \emph{\bibinfo{journal}{Journal of Physics: Condensed Matter}} \textbf{\bibinfo{volume}{26}}, \bibinfo{pages}{103202} (\bibinfo{year}{2014}).

\end{thebibliography}

\end{document}